\documentclass[twocolumn,amsmath,amssymb,floatfix]{revtex4}

\usepackage{latexsym,amsmath,amssymb,theorem}
\usepackage{amssymb} 
\usepackage{amsmath}
\usepackage{graphicx}
\usepackage{graphicx,subfigure}

\usepackage{amssymb}
\usepackage{amsmath}

\def\lsim{\mathrel{\mathstrut\smash{\ooalign{\raise2.5pt\hbox{$<$}\cr\lower2.5pt\hbox{$\sim$}}}}}
\def\gsim{\mathrel{\mathstrut\smash{\ooalign{\raise2.5pt\hbox{$>$}\cr\lower2.5pt\hbox{$\sim$}}}}}

\begin{document}

\title{Adiabatic Ekpyrosis: Scale-Invariant Curvature Perturbations from a Single Scalar Field in a Contracting Universe}

\author{Justin Khoury$^{1}$ and Paul J. Steinhardt$^2$}

\affiliation{$^1$Center for Particle Cosmology, University of Pennsylvania, Philadelphia, PA 19104 \\
  $^2$Department of Physics \& Princeton Center for Theoretical Science, Jadwin Hall, Princeton University, Princeton, New
  Jersey 08544}

                            
\begin{abstract}
The universe can be made flat and smooth by undergoing a phase of ultra-slow (ekpyrotic) contraction, a condition achievable with a single, canonical scalar field and conventional general relativity.  It has been argued, though, that generating scale-invariant density perturbations, requires at least two scalar fields and a two-step process that first produces entropy fluctuations and then converts them to curvature perturbations. In this paper, we identify a loophole in the argument and introduce an ekpyrotic model based on a single, canonical scalar field that generates nearly scale-invariant curvature fluctuations through a purely ``adiabatic mechanism" in which the background evolution is a dynamical attractor. The resulting spectrum can be slightly red with distinctive non-gaussian fluctuations. 
\end{abstract}
\maketitle 

Two mechanisms are known for making the universe flat and homogeneous in accord with cosmological observations.  The first is {\it inflation}~\cite{inf}, a period of accelerated expansion occurring after the big bang during which the equation of state $w$ is less than $-1/3$. The second is {\it ekpyrosis}~\cite{ek,Erick}, a period of ultra-slow contraction with $w \gg 1$ occurring before a bounce to an expanding phase.  Both equations of state can be obtained with a single, canonical scalar field evolving along a potential and conventional general relativity.  To obtain a nearly constant $w$, the potential must be positive and flat for inflation and steep and negative for ekpyrosis. In both cases, the scalar field and Newtonian gravitational potential obtain nearly scale-invariant fluctuations.  

In addition, the scalar field fluctuations automatically induce nearly scale-invariant curvature (density) perturbations in inflationary models.  The same is not true for ekpyrosis. It has been argued that the spectrum for the curvature perturbation $\zeta$ in ekpyrotic models with a single canonical scalar field is necessarily blue~\cite{nicolis}, inconsistent with observations of the cosmic microwave background and large-scale structure. A scale-invariant spectrum can be obtained via an ``entropic mechanism"~\cite{lehners},  but this requires at least two scalar fields and a two-step process that first produces entropy fluctuations and then converts them to $\zeta$~\cite{lehners,newek}.

Another feature of inflation is that
the cosmological background 
is a dynamical {\it attractor}, as can be 
seen from the behavior   
of
$\zeta\approx \delta a/a$ 
in the long-wavelength ($k\rightarrow 0$) limit.
Since $\zeta$, 
which measures the perturbation in the expansion
history between distant Hubble patches, approaches
a constant in this limit,
the perturbation can be absorbed through 
a spatial diffeomorphism: hence,
the background solution $a(t)$ is  an attractor. 

When compared to the known alternative cosmological models, it seemed that inflation is unique in having the combined properties of scale-invariance and dynamical attractor. For instance, the background is an attractor and $\zeta$  is constant at long wavelengths in single-field ekpyrosis~\cite{ek,nicolis}, but, as mentioned earlier, its spectrum is blue, rather than scale-invariant.
As another case in point, it is well-known that the equation for $\zeta$ in a contracting, matter-dominated universe is identical to that of inflation, and hence results in a scale-invariant spectrum~\cite{dust}; but the growing mode for $\zeta$ increases in time at long wavelengths, indicating that the background is unstable. This instability also
plagues contracting mechanisms relying on a time-dependent sound speed~\cite{piazza}.   

In this paper, though, we present a counterexample:  a model based on a single, canonical scalar field in which the contracting cosmological background is a dynamical attractor and  a scale-invariant spectrum of curvature perturbations is generated via a purely ``adiabatic mechanism.''  More precisely, the power spectrum is slightly red, $1- n_s  = {\cal O}(1 \%)$, with distinctive $k$-dependent non-gaussianity (compared to the entropic mechanism or inflation).  As with the entropic mechanism, this new adiabatic mechanism is observationally viable assuming that the bounce from contraction to expansion preserves the long-wavelength spectrum for $\zeta$.

The counterexample relies on relaxing the usual assumption that $w$ is nearly constant.  The effect can be illustrated for fairly simple scalar potentials, such as
\begin{equation}
\label{potent}
V(\phi) = V_0 (1- e^{-c \phi})\,,
\end{equation}
with $c \gg 1$. (Here and throughout, we use reduced Planck units for which the gravitational constant is $G=1/8 \pi$.)  The regime of interest is the transition when the equation of state rises rapidly from $\epsilon\equiv \frac{3}{2}(1+w) \approx 0$, where the constant term dominates, to $\epsilon \approx c^2/2$, where the negative exponential term dominates. The transition, which occurs when $\phi \sim c^{-1}\log c$ and $V$ is positive, produces an ekpyrotic phase during which scale-invariant adiabatic curvature perturbations are generated.   

The form for $V(\phi)$ is similar to inflationary examples, but here we are assuming a scenario in which the 
universe is slowly contracting just prior to the onset 
of the adiabatic mechanism. 
The transition in the equation of state occurs  over a 
tiny range $\Delta\phi\ll 1$ around $\phi = 0$,
so~(\ref{potent}) need only be a good approximation 
over this range and there is considerable freedom 
in the overall form of the potential~\cite{long}.

Note that we focus in this paper on the finite period during which the adiabatic mechanism generates perturbations presuming some pre-existing contracting phase and some subsequent bounce and reheating.  The various ways in which adiabatic mechanism can be embedded in a larger scenario 
are left for future work.

Until now, all ekpyrotic potentials required $V<0$ in order to obtain the desired properties for the equation of state: $w > 1$ and nearly constant. For example, for the negative exponential potential alone, $V(\phi) = -V_0e^{-c\phi}$,  the solution has $H = 2/c^2 t$ and $\dot{H} = -2/c^2 t^2$, where time $t \rightarrow 0^-$ corresponds to the bounce. Then, 
\begin{equation}
\epsilon \equiv \frac{3}{2} (1+w) = - \frac{\dot{H}}{H^2} \approx \, \frac{c^2}{2}\,,
\end{equation} 
a constant value much greater than unity for $c \gg 1$.  

By contrast, the scale-invariant curvature perturbations described in this paper are generated during 
the transition from the positive constant term dominating to the exponential dominating, which we term the ``transition phase" (to distinguish from the conventional ekpyrotic phase that occurs subsequently). During this transition,
we will find that the equation of state parameter $\epsilon$ grows as $1/t^2$, 
obtaining the large positive values needed to smooth and flatten the universe.
This behavior for $\epsilon$ leads to a range of wavelengths for which the spectrum
of curvature perturbations is scale invariant. 

As in inflation, the adiabatic mechanism is based on the equation of motion for the Fourier modes $\zeta_k$ with comoving wavenumbers $k$;  with a change of variable $\zeta_k(\tau)= v_k/z$, the equation of motion takes the form  
\begin{equation}
v''_k + \left(k^2 - \frac{z''}{z}\right) v_k =0\,,
\label{vpert}
\end{equation}
where primes denote derivatives with respect to conformal time $\tau$, and
\begin{equation}
z \equiv a(\tau)  \sqrt{2 \epsilon(\tau)}\,.
\end{equation}
In the ``transition phase", the scale factor is nearly constant ($a(\tau)\approx 1$) --- this is the slowly contracting background typical of ekpyrotic cosmology.
It follows that conformal time and cosmological time are approximately the same: $t\approx \tau$. Meanwhile, as we will see, the equation of state parameter satisfies 
$\epsilon\sim 1/t^2 \approx 1/\tau^2$. Therefore, $z_{\rm tran} (\tau) \sim (-\tau)^{-1}$ --- exactly as in inflation, where $\epsilon$ is nearly constant and $a(\tau) \approx 1/(-\tau)$! The spectrum is therefore scale-invariant. Note, morever, that the growing mode solution for $k\rightarrow 0$ is $\zeta = v/z \rightarrow {\rm constant}$: the ``transition phase" evolution
is a dynamical attractor.

The growing mode solution is oscillatory for $k\sqrt{| z/z''|} \gg 1 $ and constant for $k \sqrt{| z/z''|} \ll 1 $ if $z''/z>0$; the cross-over from oscillatory to constant behavior, called {\it freeze-out}, occurs when the wavelength of the mode is of order $H_{\rm freeze}^{-1} \equiv \sqrt{| z/z''|}$, the {\it freeze-out horizon} radius.  The notation is chosen to remind the reader that the freeze-out horizon, the relevant scale for analyzing perturbations, is of order the Hubble horizon $H^{-1}$ for constant $\epsilon$; in this paper, though, it will be important to distinguish the two horizons.

The conditions $\epsilon \sim 1/t^2$ arises naturally as a background solution for simple potentials, such as~(\ref{potent}).
As usual in ekpyrotic cosmology, the background scalar field evolution is insensitive to the slowly-contracting
background and is therefore driven by $V(\phi)$:
\begin{equation}
\ddot{\phi} \approx -c V_0e^{-c\phi}\,.
\label{phieom}
\end{equation}
We will check {\it a posteriori} that gravity can indeed be neglected in this equation. 
Note that the evolution of $\phi$ is oblivious to the constant term in
$V(\phi)$. The solution is therefore of the standard ekpyrotic form: 
\begin{equation}
\phi(t) \approx \frac{2}{c}\log\left(\sqrt{\frac{V_0}{2}}c |t|\right)\,.
\label{phisol}
\end{equation}
Substituting into $\dot{H} = -\dot{\phi}^2/2$ gives $\dot{H} \approx -2/c^2t^2$, 
which can be immediately integrated: $H(t) = 2/c^2t + H_0$. At sufficiently early times,
$H$ is nearly constant, with the constant $H_0$ fixed by the Friedmann equation:
$3H_0^2 \approx  \dot{\phi}^2/2 + V(\phi)  = V_0$. This defines the transition phase.
In other words, the solution~(\ref{phisol}) is such that the kinetic energy nearly cancels
the exponential term in~(\ref{potent}), leaving the constant $V_0$ term as the dominant contribution to $H$:
\begin{equation}
H_{\rm tran}(t) \approx \frac{2}{c^2t} -\sqrt{\frac{V_0}{3}}\approx  -\sqrt{\frac{V_0}{3}} \,.
\label{H}
\end{equation}
The transition phase ends when the constant term no longer dominates, which can be read off from~(\ref{H}):
\begin{equation}
t_{\rm end-tran} \equiv t_{\rm beg-ek} = -\frac{2}{c^2}\sqrt{\frac{3}{V_0}}\,.
\label{tend}
\end{equation}
For $ t > t_{\rm end-tran}$, we have $H(t)\approx 2/c^2t$, and the solution reduces to a standard ekpyrotic scaling phase. 

Meanwhile, the equation of state follows from~(\ref{H}):
\begin{equation}
\epsilon = -\frac{\dot{H}}{H^2} = \frac{6}{c^2V_0}\frac{1}{(t+t_{\rm end-tran})^2}\,.
\label{epsapprox}
\end{equation}
Deep in the transition phase, $|t|\gg |t_{\rm end-tran}|$, we have $\epsilon\sim 1/t^2$, as desired to generate a
scale-invariant spectrum for $\zeta$. For $|t| \ll |t_{\rm end-tran}|$, however, we have $\epsilon\approx c^2/2$,
indicating again an ekpyrotic scaling phase.

Let us check the validity of assuming $H\dot{\phi} \ll c V_0e^{-c\phi}$ in~(\ref{phieom}).
Substituting $H\approx -\sqrt{V_0/3}$ and the solution~(\ref{phisol}), it is easily seen that this approximation is
consistent for
\begin{equation}
t > t_{\rm beg-tran} \equiv - \sqrt{\frac{3}{V_0}}\,.
\label{tbegin}
\end{equation}
The transition is shorter than a Hubble time, and hence the scale factor, $a_{\rm tran}(t) \approx 1 - \sqrt{V_0/3}\;t$ is approximately constant
throughout --- the universe is nearly static.

Although $H$ is nearly constant during the transition phase, this is emphatically {\it not} a contracting de Sitter universe. There are many equivalent ways to see this.
First, as was already mentioned, the transition from $\epsilon\approx 0$ to $\epsilon \approx c^2/2$ occurs within a Hubble time, and hence $a(t)\approx 1$.
Moreover, the rate of change of $\epsilon$ is never small:
$\eta \equiv H^{-1}{\rm d}\ln \epsilon/{\rm d}t$ ranges from ${\cal O}(1)$ to $c^2\gg 1$.

The transition phase is followed by a standard scaling ekpyrotic  phase in which $\epsilon\approx c^2/2$. For $|t|\ll  |t_{\rm end-tran}|$,
the background is therefore slowly-contracting with~\cite{ek} 
\begin{equation}
a_{\rm scaling} (t) \sim (-t)^{2/c^2}\;;\qquad H_{\rm scaling}(t) \approx  \frac{2}{c^2t} \,.
\label{ekstd}
\end{equation}
This standard ekpyrotic phase is part of the story as well: it enables the Hubble horizon to shrink while preserving
the scale-invariant spectrum such that the modes are well-outside the Hubble horizon before the bounce.

As for $\zeta$, we have from~(\ref{epsapprox}) and $a(t)\approx 1$ that
\begin{equation}
z_{\rm tran}(t)  = a(t)\sqrt{2\epsilon(t)} \approx \frac{2\sqrt{3}}{c\sqrt{V_0}(-t)}\,
\end{equation}
within the transition phase when $|t|\gg |t_{\rm end-tran}|$.
Thus, $z''/z\approx \ddot{z}/z\approx 2/t^2$, and the solution to~(\ref{vpert}) for the mode functions, with the adiabatic vacuum choice, is
\begin{equation}
v_k = \frac{e^{-ikt}}{\sqrt{2k}}\left(1-\frac{i}{kt}\right)\,.
\end{equation}
These mode functions are identical to those in de Sitter space, except that their amplification
relies here on a rapidly-changing $\epsilon(t)$, not a rapidly-changing $a(t)$. The long-wavelength amplitude for the curvature perturbation, $\zeta_k = v_k/z$, is therefore
scale-invariant:
\begin{equation}
k^{3/2}|\zeta_k|  = c\sqrt{\frac{V_0}{24}}\,.
\label{zetafin}
\end{equation}
The range of these modes is set by the duration of the transition phase:
$k_{\rm max}/k_{\rm min} = t_{\rm beg-tran}/t_{\rm end-tran} = c^2/2$.  

The transition phase is an attractor. Following~\cite{nicolis}, we can compute perturbations to $g_{ij}$ in synchronous gauge and show that 
they all go to zero as ${\cal O}(k|t|)$ or as $\log(t/t_{\rm end-tran})$, except for the $\zeta$ term which goes to a constant as usual. Thus the metric approaches its unperturbed form. Similarly, $\delta\phi ={\cal O}(k|t|)$.

In terms of local observables, the 3-curvature scalar tends to a constant, $R^{(3)}\sim k^2\zeta/a^2$, as usual.
Meanwhile, the extrinsic curvature tensor, $K_{ij} = \dot{g}_{ij}/2$, receives a growing contribution from the log term in $g_{ij}$:
\begin{equation}
\frac{\delta K}{\bar{K}} \sim \frac{k^2}{V_0}\frac{\zeta}{c^2\sqrt{V_0}t}\,.
\label{delK}
\end{equation}
However, recall that the background itself evolves away from de Sitter as $H(t) = -\sqrt{V_0/3} + 2/c^2t$. Thinking of this as
a correction over $\bar{K} = -\sqrt{3V_0}$, we find
\begin{equation}
\frac{\Delta H}{H} \sim \frac{1}{c^2\sqrt{V_0}t}\,.
\end{equation}
That is, the perturbation~(\ref{delK}) behaves identically to the background. Even more convincing is the observation that~(\ref{delK}) is less singular than $\delta H/H \sim 1/t^2$ resulting from
a constant time shift of the background solution. 

To incorporate our mechanism in an observationally acceptable model, there are three further considerations: 

\noindent {\it Hubble horizon exit}: Although modes are amplified during the transition phase,
they have not yet exited the actual Hubble horizon. Since both $a(t)$ and $H(t)$ are
constant throughout this phase, modes remain within the Hubble horizon. Mode amplification
nevertheless occurs because the freeze-out horizon radius, $H_{\rm freeze}^{-1} \equiv \sqrt{| z/z''|}$,
shrinks, thanks to the rapidly-changing $\epsilon(t)$. 

The exit of modes outside the actual Hubble horizon is achieved during the subsequent
scaling ekpyrotic phase, since $|H|^{-1}\sim |t|$ shrinks in this period. But since $a(t)\sim (-t)^{2/c^2}$
with $c\gg 1$ during this phase, the background is very slowly contracting, and the scale-invariant
spectrum is preserved in the process. In other words, instead of $\zeta_k \sim k^{-1/2}z^{-1}$, as is the case when the wavelength is first well below $H_{\rm freeze}^{-1}$,  the appropriate initial conditions
for the mode functions at the onset of the ekpyrotic phase is modified: $\zeta_k \sim k^{-3/2}z^{-1}$. Matching at horizon crossing
gives $k^{3/2}\zeta_k \sim k^{-1/(1-\epsilon)}$, which is nearly scale-invariant for $\epsilon \approx c^2/2\gg 1$. 

\begin{figure} 
   \centering
   \includegraphics[width=3.0in]{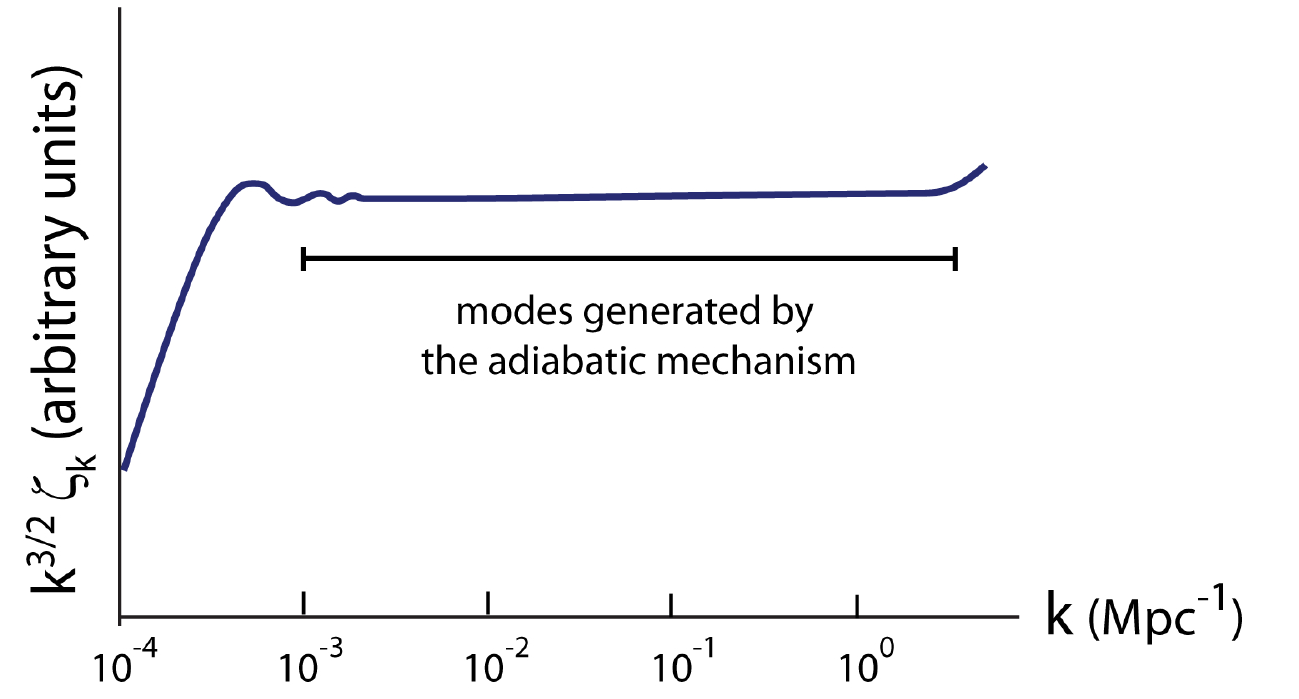}
   \caption{Numerical computation of the perturbation amplitude $k^{3/2} \zeta$ vs. $k$ generated by the adiabatic mechanism.  The behavior of modes with larger and smaller $k$ depends on the larger scenario in which the mechanism is embedded and beyond the consideration of this paper.}
   \label{zetak}
\end{figure}

These conclusions are borne out by numerical analysis. Using $z = 2c(-t)^{2/c^2}/(1 + c^2H_0t)$ to cover
the transition and scaling phases, we integrate~(\ref{vpert}) with $c=140$ and $|H_0| = 5\cdot 10^{-4}$
over the interval $-2.5|H_0|^{-1} < t <  -5\cdot 10^{-10} |H_0|^{-1}$, for the range of modes
$0.02|H_0| < k < 2\cdot 10^4 |H_0|$. The shortest-wavelength mode is therefore
barely outside the Hubble radius by the end of the integration, which occurs deep in the
ekpyrotic scaling phase. Figure~\ref{zetak} shows the resulting spectrum, with
range of scale-invariant modes spanning a factor of $10^4 \sim c^2$. 

We can immediately derive constraints on $V_0$ and $c$. Since the scale-invariant modes are on scales smaller than $|H|^{-1} \approx \sqrt{3/V_0}$,
a necessary condition is that the comoving scale $1/\sqrt{V_0}$ encompasses the entire observable universe. If the value of $H$ at the
end of the ekpyrotic phase, $H_{\rm ek-end}$, is comparable to that at the onset of the expanding, radiation-dominated phase, 
we must have $H_{\rm ek-end}/\sqrt{V_0} \;\gsim\; a_{\rm ek-end}H_{\rm ek-end}/(a_{\rm today}H_{\rm today}) \sim \sqrt{H_{\rm ek-end}/H_{\rm today}}$,
where we have assumed a radiation-dominated evolution until the present epoch for simplicity. In other words,
\begin{equation}
\sqrt{V_0} \;\lsim \; \sqrt{H_{\rm ek-end}H_{\rm today}} \approx 10^{-30} H_{\rm ek-end}^{1/2}\,.
\label{V0cons}
\end{equation}
For Grand-Unified (GUT) reheating scale, $H_{\rm ek-end} \sim 10^{12}$~GeV, and the above condition is satisfied for $V_0\sim {\rm TeV}^4$.
Setting $\zeta\sim 10^{-5}$ on large scales then requires from~(\ref{zetafin}) that $c = 10^{28}$. For electroweak (EW) reheating,
$H_{\rm ek-end} \sim {\rm meV}$, we similarly get $V_0 \sim {\rm MeV}^4$ and $c = 10^{40}$. 

Although we have focused on pure exponential potentials, for simplicity, the exponentially large values of $c$ required can be achieved effectively in the
Conlon-Quevedo potential~\cite{CQ}, $V(\phi) \sim \exp(-\gamma\phi^{4/3})$, for  large $\phi$.

\noindent {\it Spectral Index}: Since the values of $c$ of interest are exponentially large, the departures from
scale-invariance are thus far unobservable. However we can generate a small red tilt, as favored by
observations~\cite{wmap5}, by generalizing~(\ref{potent}) to allow for a slowly-varying exponent $c(\phi)$:
\begin{equation}
V(\phi) = V_0 + U(\phi) = V_0 \left(1- e^{-\int {\rm d}\phi \; c(\phi)}\right)\,.
\label{tiltpot}
\end{equation}
In the transition-phase approximation, the Hubble parameter is constant, $H\approx -\sqrt{V_0/3}$, whereas
$U(\phi)$ cancels against the kinetic energy of $\phi$. Therefore,
\begin{equation}
\epsilon  \equiv -\frac{\dot{H}}{H^2} = \frac{3\dot{\phi}^2}{2V_0} \approx -\frac{3U(\phi)}{V_0}\,.
\end{equation}
Since $a(t)\approx 1$ as before, it follows that $z\sim \sqrt{-U(\phi)}$. Using the evolution equation $\dot{\phi} = -\sqrt{-2U(\phi)}$, it is straightforward to show that
$z''/z\approx \ddot{z}/z \approx -U_{,\phi\phi}$,
where
$U(\phi)$ can be shown to satisfy $U_{,\phi\phi} \approx  -(2/t^2)(1+ 3c_{,\phi}/c^2)$~\cite{lehners}. Substituting in~(\ref{vpert}),
the resulting tilt is
\begin{equation}
n_s - 1  = -4\frac{c_{,\phi}}{c^2}\,.
\end{equation}
Since $\phi$ is decreasing in our solution, the spectral tilt will be slightly red if $c_{,\phi}> 0$.
For instance, if $c(\phi)$ changes smoothly by ${\cal O}(c)$ during the transition, then $n_s - 1  \approx -4 (\Delta c)/(c^2\Delta\phi)\approx -4/(c\Delta\phi)$.
Using~(\ref{phisol}), we obtain
\begin{equation}
n_s - 1 \approx -\frac{2}{\log(t_{\rm beg-tran}/t_{\rm end-tran})} = -\frac{2}{\log(c^2/2)}\,.
\end{equation}
For $10^{40} > c  > 10^{28}$, ranging from EW to GUT-scale reheating, we obtain
$n_s \approx 0.98$. Since we dropped ${\cal O}(1)$ factors, the generic answer is
$1-n_s  \approx {\rm few} \;\%$, in good agreement with observations~\cite{wmap5}.

\noindent {\it Non-Gaussianity}: 
As with the entropic mechanism~\cite{intuitive}, a steep non-linear potential generates large non-gaussianity compared to simple inflationary models. In this case, the steepness increases during the transition phase, so modes that freeze out later have larger non-gaussianity; {\it i.e.}, the non-gaussianity is $k$-dependent.  The precise calculation is lengthy and will be presented in~\cite{long}. 
The result is $\langle \zeta_{k_1}\zeta_{k_2}\zeta_{k_3}\rangle = 4 (2\pi)^3 \delta(\vec{k}_1+\vec{k}_2+\vec{k}_3) k^6|\zeta_k|^4 {\cal A}/\prod_j k_j^3$ with
\begin{equation}
{\cal A} \approx -\frac{3k^2}{4c^2V_0}\left(\sum_ik_i^3 - \sum_{i\neq j}k_ik_j^2+2k_1k_2k_3\right)\,,
\label{3amp}
\end{equation}
where $k \equiv k_1+k_2+k_3$. This amplitude peaks for equilateral configurations ($k_1\approx k_2\approx k_3$) and vanishes in the squeezed or local limit ($k_3\ll k_1\approx k_2$)~\cite{long}. As usual, the magnitude of ${\cal A}$ can be characterized by $f_{\rm NL}^{\rm equil} \equiv 30 {\cal A}_{k_i = k/3}/k^3$, which in our case gives
\begin{equation}
f_{\rm NL}^{\rm equil}  = \frac{5k^2}{6c^2V_0}\,.
\label{fNLc}
\end{equation}
Note that $f_{\rm NL}^{\rm equil} $ increases with $k$ and is of order $c^2$ for the shortest-wavelength mode generated during the transition phase ($k_{\rm max} = 1/t_{\rm end-tran}\sim c^2/\sqrt{V_0}$). Given that exponentially large values of $c$ are needed to have $\zeta \sim 10^{-5}$, there is the danger that the non-gaussianity grows unacceptably large for some range of $k$. (Ref.~\cite{mukhanovlinde} recently pointed out additional problems with the pure exponential model.) The problems arise because the ekpyrotic phase with                                                                       
large $c$ is maintained longer than needed.  The simple solution is to alter the pure exponential potential so as to  terminate the transition                                                                       
phase before the modes with unacceptably large $f_{\rm NL}^{\rm equil}$ are generated; for example, if $c(\phi)$ decreases smoothly to some $b\ll c$ after the transition phase has generated an observationally acceptable range of scale-invariant fluctuations. Current observations require $\zeta \sim 10^{-5}$ over a range of $k$ spanning a factor of $10^3$, which can be  achieved while keeping the non-gaussianity acceptably small. Terminating the transition phase results in the perturbation amplitude being suppressed on smaller scales: the spectrum tilts strongly to the red and then flattens out at an exponentially smaller amplitude with an acceptable non-gaussianity ($f_{\rm NL} \zeta \ll 1$) throughout. The perturbative regime is consistently valid, both classically and quantum mechanically, at all times in this model~\cite{long}.


We close by noting that the generation of a scale-invariant spectrum of adiabatic curvature perturbations over a range of $k$ spanning $10^3$ or so was once a theoretical speculation and, after the introduction of inflationary cosmology, a theoretical prediction.  Today it is a rigorous criterion for a viable cosmological model, given the observations of the microwave background and large-scale structure.  What has been learned from the ekpyrotic model is that the spectrum is not unique to inflation. With this paper, we have introduced an adiabatic ekpyrotic mechanism that generates scale-invariant  adiabatic perturbations, within a background that is a dynamical attractor. This mechanism also predicts non-gaussianity and a spectrum of gravitational waves that is observationally distinguishable from inflation~\cite{long}.

{\it Acknowledgments:} We thank A.~Nicolis, A.~Tolley and N.~Turok for discussions and insights.
This work was supported in part by the US Department of Energy grant DE-FG02-91ER40671 (PJS),
by the University of Pennsylvania and by NSERC of Canada (JK).

\end{document}